\documentclass[final,1p,times]{elsarticle}
\usepackage{amssymb}
\usepackage[T1]{fontenc}
\usepackage[latin1]{inputenc}
\usepackage[font=small,labelfont=bf]{caption}
\usepackage{lineno,hyperref}
\usepackage{graphicx}
\modulolinenumbers[5]

\journal{Nuclear Instruments and Methods in Physics Research Section A}

\begin{document}

\begin{frontmatter}

\title{A method to define the energy threshold depending on noise level for rare event searches}

\author{M. Mancuso\footnote{Corresponding author.}}

\author{A. Bento\footnote{Also at LIBPhys, Departamento de Fisica, Universidade de Coimbra, P3004 516 Coimbra, Portugal.}}
\author{N. Ferreiro Iachellini}
\author{D. Hauff}
\author{F. Petricca}
\author{F. Pr\"obst}
\author{J. Rothe}
\author{R. Strauss}

\address{Max-Planck-Institut f\"ur Physik, F\"ohringer Ring 6, M\"unchen, Germany}

\begin{abstract}
Solid state detectors and low-temperature calorimeters are widely employed in rare event searches, because of their excellent sensitivity and consequent low energy threshold. A common procedure to establish the energy threshold is to define the trigger level at a fixed number of baseline standard deviations (sigmas) above the baseline level, typically 3$\sigma$ or 5$\sigma$. This is not an ideal option when the threshold becomes a critical parameter for the detectors, requiring an optimised definition of the trigger. Recorded events with a small amplitude-to-noise ratio likely survive the selection criteria of the analysis chain contributing to the experiment background. We present a method to quantify the lowest trigger threshold achievable as a function of the acceptable amount of noise events triggered for the physics case under investigation. We then apply this novel method to existing experimental and simulated data to validate the model we presented. 

\end{abstract}

\end{frontmatter}


\section{Introduction}
\label{sec:intro}
There is a widespread tendency, among rare event searches, to continuously lower the energy range of interest. Among those using cryogenic and solid-state detectors it is certainly the case for low-mass dark matter direct detection experiments \cite{cresst}\cite{edelweiss}\cite{superCDMS} and coherent neutrino nucleus scattering~\cite{nunu}\cite{miner}\cite{coherent}\cite{RICOCHET}\cite{TEXONO}\cite{Heusser2015}. For energy reconstruction close to the threshold, which is critical for the sensitivity goal of these experiment, a thorough understanding of the possible role of noise contributions is essential.\\
In this paper, we investigate how noise affects the measured energy spectrum in the vicinity of the trigger threshold. Random fluctuations in noise become an irreducible background contribution close to the threshold because pulse shape selection parameters are rendered insensitive in a low signal-to-noise ratio regime. For most cases, the trigger level is defined in terms of a multiple number of the standard deviation $\sigma$ of the baseline noise distribution above the average baseline level. We propose instead to set the threshold according to the total rate of accepted noise triggers.This approach keeps the background contribution of triggering in the noise low enough to be considered negligible with respect to the expected signal of interest. The aim of this work is to evaluate and analyze the contribution of triggers in noise, as a function of the threshold value, to the energy spectrum.\\
In sections~\ref{sec:filter} and \ref{sec:trigger}, we describe the data processing and the trigger algorithm used. We adopted the optimum filter~\cite{gatti} concept to evaluate the signal amplitude of a pulse (estimator of the energy deposition). This data treatment provides the best signal-to-noise ratio, and data filtered in this way are therefore optimised for triggering. In sec.~\ref{sec:threshold}, we discuss the ideal case of noise uncorrelated in time and derive the analytical description of noise trigger rate, testing it with a simulation. In sec.~\ref{sec:exp}, we quantify the noise trigger rate for a specific experimental set-up in order to illustrate an optimization procedure for assessing the optimal energy threshold, depending on the application. In sec.~\ref{sec:corr} we explore a realistic case when noise correlation is present.\\

\section{Optimal data filtering}
\label{sec:filter}
\begin{figure}[h!]%
\begin{center}
\includegraphics[width=12cm]{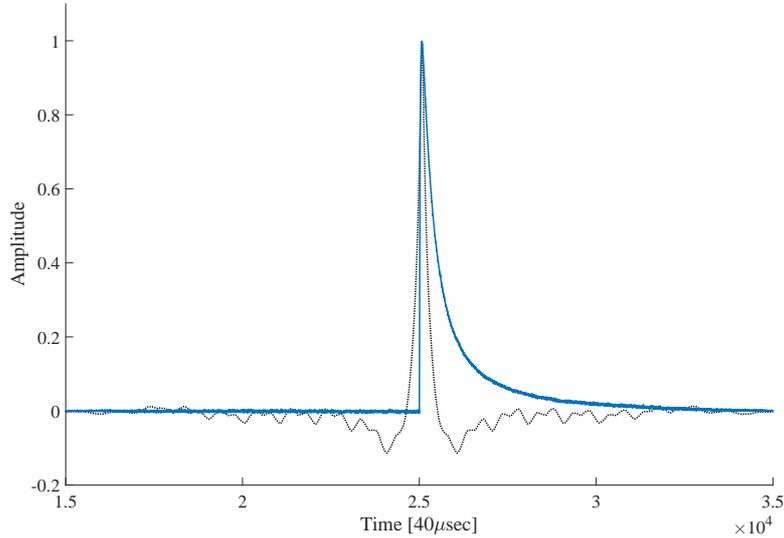}%
\caption{Solid line: A typical thermal pulse of the detector used as example in sec.~\ref{sec:exp} with amplitude normalized to one. Dotted line: The same pulse filtered with eq.~\ref{eq:gatti}. The filtered response pulse is symmetric around the maximum and the lobes present in the tails are due to deformations of the signal after filtering.}%
\label{Fig:Pulse}%
\end{center}	
\end{figure}

The first step of the trigger process is data filtering. Signal filtering is a powerful tool to reduce noise contribution in a signal, leading to an improved signal-to-noise ratio, and consequently a better resolution and lower energy threshold. In the case of a known detector response, matched filters provide the best results. The transfer function of a matched filter is obtained via maximization of the filter's response with respect to quantities of interest, such as signal amplitude, signal shape parameter, or signal-to-noise ratio. Since their definition follows an optimization problem, matched filters are usually called optimum filters.\\
According to~\cite{gatti}, there is a unique transfer function $H(\omega)$ which maximises the ratio between the filtered pulse amplitude at time $\tau_M$ and the noise RMS after filtering:
\begin{equation}
H(\omega)=K\frac{S^*(\omega)}{\mathcal{N}(\omega)}e^{-j\omega \tau_M},
 \label{eq:gatti}
\end{equation}
where $S^*(\omega)$ is the complex conjugate of the Fourier transform of the pulse shape function $s(t)$, and $K$ is a constant. This transfer function $H(\omega)$ is solely built using the noise power spectrum $\mathcal{N}(\omega)$ and the pulse shape function.\\
For our illustration, we used data from a particular cryogenic calorimeter, but in general all the following considerations apply to any calorimetric detector, namely a detector that provides a measurement of the total energy deposition. In such a detector, the signal amplitude is usually the parameter measure from which the energy spectrum is obtained. The signal shape of the detector used in this work is shown in fig.~\ref{Fig:Pulse}~(solid line). The transformed signal in the time domain is illustrated in fig.~\ref{Fig:Pulse}~(dotted line).

\section{Trigger algorithm}
\label{sec:trigger}

Many applications require a low energy threshold and it is therefore desirable to tag events as close as possible to the noise level. Compared to the raw data from the output of the detector, the output of the optimum filter shows a larger signal-to-noise ratio, allowing for a lower threshold. However, the numerical filtering of the entire stream of raw data is not straightforward, and the computational procedure used for this purpose is described in \cite{lower}. The optimisation of the trigger requires a dedicated step in the data analysis. The algorithm requires the transfer function of the filter as an input, implying that the expected shape of signal and noise power spectra must be known. In order to obtain these, a short data processing step is needed before the actual data triggering.\\
In the following, we define a simple trigger algorithm to precisely flag the arrival time of the transformed signal in the time domain. We aimed for an analytical description of the number of noise-induced triggers~(sec.~\ref{sec:threshold}) as a function of the threshold value. Such a description (model) is indeed determined by the choice of the trigger algorithm.\\
The response function of a pulse for the filter used in this work is symmetric in time with respect to the position of the maximum (fig.~\ref{Fig:Pulse}). The transformation produces lobes in the tails of the signal in the presence of periodic noise populating the bandwidth of the pulses. The lobes are proportional to the pulse amplitude and they can themselves exceed the threshold in the vicinity of a large signal. Due to these effects, an algorithm to detect only the global maximum of a pulse is required. We defined a time window that must be wide enough to contain the full waveform of the filtered pulse. As soon as the signal exceeds the threshold value, we searched for an absolute maximum, and the time window was then centred around the sample above threshold. The maximum found can be the pulse maximum or a lobe in the tails. To resolve such cases, the algorithm searches for an absolute maximum inside a new window centred on the current maximum. If a different absolute maximum is found, the search window centres on it and the procedure is iterated until the maximum of the triggered events is found.\\
An appropriate adjustment of the dead time must be applied for this triggering method. As a triggered event has to be separated by more than a half-window length from another trigger, the whole time window length has to be considered dead time for any energy smaller than the triggered pulse. The resulting energy-dependent dead time can be easily obtained from the final energy spectrum and included in the trigger efficiency, which then depends not only on the rate but also the shape of the measured spectrum. This simplistic dead time calculation is valid for data-sets where the rate is low enough to neglect the probability of three or more events that pile up, certainly the case in most rare event searches. 

\section{Analytical description of the noise trigger rate}
\label{sec:threshold}

\begin{figure}[h!]%
\begin{center}
\includegraphics[width=1\textwidth]{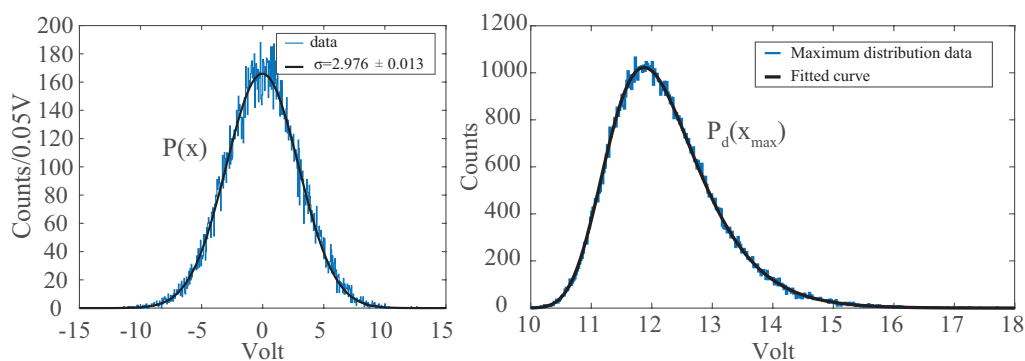}%
\caption{Left: the sample distribution of a simulated noise window fitted with a Gaussian function. Right: Distribution of the maxima of 40000 baseline windows of 25000 samples, the black curve is the result of the fit according to eq.\ref{eq:stats1}.}%
\label{Fig:noise}%
\end{center}	
\end{figure}

To evaluate the number of noise triggers for the algorithm illustrated above, given a particular noise spectrum, we have to describe the probability that in a time window containing only noise, the maximum value of the samples\footnote{We use "`sample"' to define a single digitised value of the detector output, and "`window"' for a set of sequential samples that describe a waveform.} exceeds the threshold value. This requires firstly to describe the distribution probability $P(x)$ of filtered noise samples, and secondly to define the length of the trigger window, $d$. Once these two quantities are chosen, it is possible to describe the distribution of the maxima for a set of filtered baseline windows using combinatorics. In other words, we looked for the probability function $P_{d}\left(x_{max}\right)$ which describes the probability to find a maximum sample value equal to $x_{max}$ in a noise window of length $d$, after filtering.\\
The simple calculation for the ideal case with statistically independent noise samples follows below. The more realistic case of correlated samples can then be recovered by an independent fit of the number of statistically independent samples $d_{eff}$ in the trigger window, see sec.~\ref{sec:corr}.\\ 
Assuming that each sample follows the distribution $P(x)$ and that they are statistically independent, the joint probability of one sample being equal to $x_{max}$, and all others being smaller, can be expressed with a binomial distribution:
\begin{equation}
P_{d}\left(x_{max}\right)=\frac{d!}{1!\left(d-1\right)!}\left(P\left(x_{max}\right)\right)\left(\int^{x_{max}}_{-\infty}P(x)dx\right)^{d-1}.
\label{eq:stats}
\end{equation}
In many cases $P(x)$ can be described by a Gaussian function. By using the Gaussian error function $erf(x)$ to describe the integral of the normal distribution $P(x)$, the probability distribution $P_{d}\left(x_{max}\right)$ can be formulated as:
\begin{equation}
P_{d}\left(x_{max}\right)=\frac{d}{\sqrt{2\cdot\pi}\cdot\sigma}\cdot \left(e^{-\left(\frac{x_{max}}{\sqrt{2}\sigma}\right)^2}\right)\cdot\left(\frac{1}{2}+\frac{{erf}\left(x_{max}/(\sqrt{2}\sigma)\right)}{2}\right)^{d-1}.
\label{eq:stats1}
\end{equation}
The total Noise Trigger Rate (NTR) above threshold, expressed in units/(kg$\times$day), is then
\begin{equation}
NTR\left(x_{th}\right) =\frac{1}{t_{win}\cdot m_{det}}\int^{\infty}_{x_{th}}P_{d}\left(x_{max}\right)dx_{max},
\label{eq:stats2}
\end{equation}
where $t_{win}$ is the trigger time window, $x_{th}$ is the threshold value, and $m_{det}$ is the detector mass. Signal and background processes in rare event searches typically scale with detector mass (units/(kg$\times$day)), so for a meaningful comparison the NTR needs to be converted to this unit.\\
From eq.~\ref{eq:stats2} it is possible to determine the value of the threshold, $x_{th}$, according to a desired accepted rate of triggers in noise. This approach provides a tool to define the value of the energy threshold adjusted to the acceptable rate of misidentified noise events according to the application. Furthermore, eq.~\ref{eq:stats} also describes the energy distribution of this background, which can then be accounted for in background models.\\
To validate the model, we simulate a set of 40000 windows of white noise, all with the same characteristics: average~=~0~V, $\sigma$~=~3~V, and 25000 samples. The distribution of the samples ($P(x)$) is shown in Fig.~\ref{Fig:noise}(left). For each noise window we compute the maximum value of the samples. A fit of the spectrum of the baselines maxima with eq.~\ref{eq:stats1}, where the parameters $\sigma$ and $d$ are left free, is displayed in fig.~\ref{Fig:noise}(right). The result of the fit (tab.~\ref{tablefit}) is in agreement with the input parameters within the statistical error.

\begin{table}%
\centering
\begin{tabular}{ccc}
\hline
Parameter & input value & fit result\\
\hline
$\sigma$ & 3 V & $3$ $\pm$ $0.003$ V\\
$d$ & 25000 & $25010$ $\pm$ $490$\\
\hline
\\
\end{tabular}
\captionof{table}{Fit result for the distribution of the baseline maxima from simulation.\\ }
\label{tablefit}
\end{table}

\section{Proof of principle application}
\label{sec:exp}
\begin{figure}[htbp]
\centering 
\includegraphics[width=.95\textwidth]{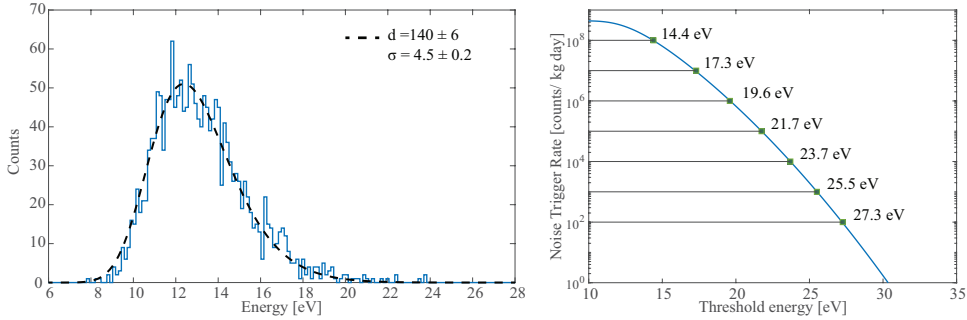}
\caption{Left: The solid represents the histogram of the maxima found in $\sim$1400 baselines available within the test measurement, the dashed line corresponds to the fit result using eq.~\ref{eq:stats1}. Right: Noise Trigger Rate as function of threshold, some representative values are displayed (see text).}
\label{fig:maxdistreal}
\end{figure}

We now show the result of the threshold assignment method applied to an interesting real case. We use the data acquired with a gram-scale cryogenic detector prototype developed for the $\nu$-cleus experiment~\cite{nunu}. This project aims at the detection of low-energy nuclear recoils from coherent neutrino-nucleus scattering~\cite{nunu}~\cite{gram} and is also used to derive a limit on MeV-scale Dark Matter~\cite{MeV}. The sensitivity to these processes is enhanced by a low energy threshold~\cite{lowenergy}. This measurement provides a suitable test bench for the method proposed. The detector consists of a gram-scale calorimeter, an Al$_2$O$_3$ cube of 5$\times$5$\times$5~mm$^3$with a mass of 0.5~g. The detector was operated in a dilution cryostat at the Max-Planck-Institut (MPI) for Physics in Munich, Germany. A more detailed description of the measurement can be found in~\cite{gram}\\
We apply the procedure described in sec.~\ref{sec:trigger}-\ref{sec:threshold} on the data. The reference pulse used to construct the filter is shown in fig.~\ref{Fig:Pulse}. The data are continuously filtered in the frequency domain with the optimum filter described in sec.~\ref{sec:filter}. The filtered waveforms are reconstructed in the time domain. A set of baseline windows are selected using criteria based on the RMS of the unknown waveform. For consistency with the selection of baselines, the same RMS cut used for the selection of the baselines is applied to the baseline samples before the onset of the pulse to ensure the same noise conditions.\\
A first simple trigger without optimisation is used for calibration and performance studies. This analysis was made independently from the one in~\cite{gram}. The baseline fluctuation after filtering corresponds to $\sigma~=~(3.5\pm0.4)~$eV. The calibration factor is obtained using the 5.9~keV X-rays of the $^{55}$Fe calibration source.\\
The first data pre-processing provides the set of empty baselines suitable to evaluate a trigger threshold for the measurement considered. Noise time windows (empty baselines) of 400 ms, which is the length of the trigger sliding window, are selected from the data to build the distribution of the maxima. The fit of the distribution using eq.~\ref{eq:stats1} is shown in fig.~\ref{fig:maxdistreal}~(left). Similar to the simulated case, we let the parameters $\sigma$ and $d$ free. The result of the fit can be used to compute the noise trigger rate (NTR), eq.~\ref{eq:stats2}, as a function of the trigger threshold (fig.~\ref{fig:maxdistreal}~(right)). Despite the distribution being well-fitted ($\tilde{\chi}^2$~=~0.96), the parameter $d$~=~208~$\pm$~6 is far from the expected value of 10000 and $\sigma$~=~4.5 $\pm$~0.2~eV differs significantly from the baseline fluctuation $\sigma~=~3.5~$eV.\\
As $d$ is the number of statistically independent samples in a trigger window, the result suggests that after filtering, the data still have large components of correlated noise, unlike in the simulated data where only uncorrelated noise is considered. It is therefore expected that the number of independent samples is smaller than the frequency sampling multiplied by the length of the window. A dedicated discussion follows in the next section sec.~\ref{sec:corr}.
The detector baseline fluctuation of $\sigma~=~3.5~$eV allows for the identification of $\sim$20 eV signals with high efficiency \cite{gram}. However, with this measurement condition the noise trigger rate with a threshold of $\sim$20 eV results in $\sim$10$^6$ counts/(kg day). In fig.~\ref{fig:maxdistreal}~(right) different threshold values are highlighted corresponding to noise trigger rate from 10$^{2}$ to 10$^8$ counts/(kg day). The measured background rate of 10$^8$ counts/(kg day) in the energy region below 1 keV \cite{MeV} cannot ensure a baselines sample free from pulses, a condition which would be better fulfilled in an underground measurement. The distribution of maxima is therefore affected by undetected pulses which extend the tail to higher energies and explain the discrepancy in the $\sigma$ value obtained by the fit of the baseline distribution and the fit of the maxima distribution. A value of $d$ smaller than the expected one also extends the tail of the noise trigger distribution to higher energy (eq.~\ref{eq:stats1}), explaining why this detector, triggered at a threshold of $\sim$20 eV ($>5\cdot \sigma$ of the baseline distribution), still has a very high NTR. By suppressing the rate, using a shielded or underground set-up, the probability of small pulses hidden in the noise baselines window would be lower and thus a further reduction of noise leakage above threshold is expected.\\ 
In order to set the ideal threshold value for the measurement under consideration, we compare the expected signal rate with the background rate due to noise triggers. Using the value computed with the fit, we can calculate the amount of noise triggers in the test measurement and the threshold value which ensures a negligible contribution of triggers in noise for that particular detector application.\\ 

\section{Considerations on correlated noise}
\label{sec:corr}
\begin{figure}[htbp]
\centering 
\includegraphics[width=.95\textwidth]{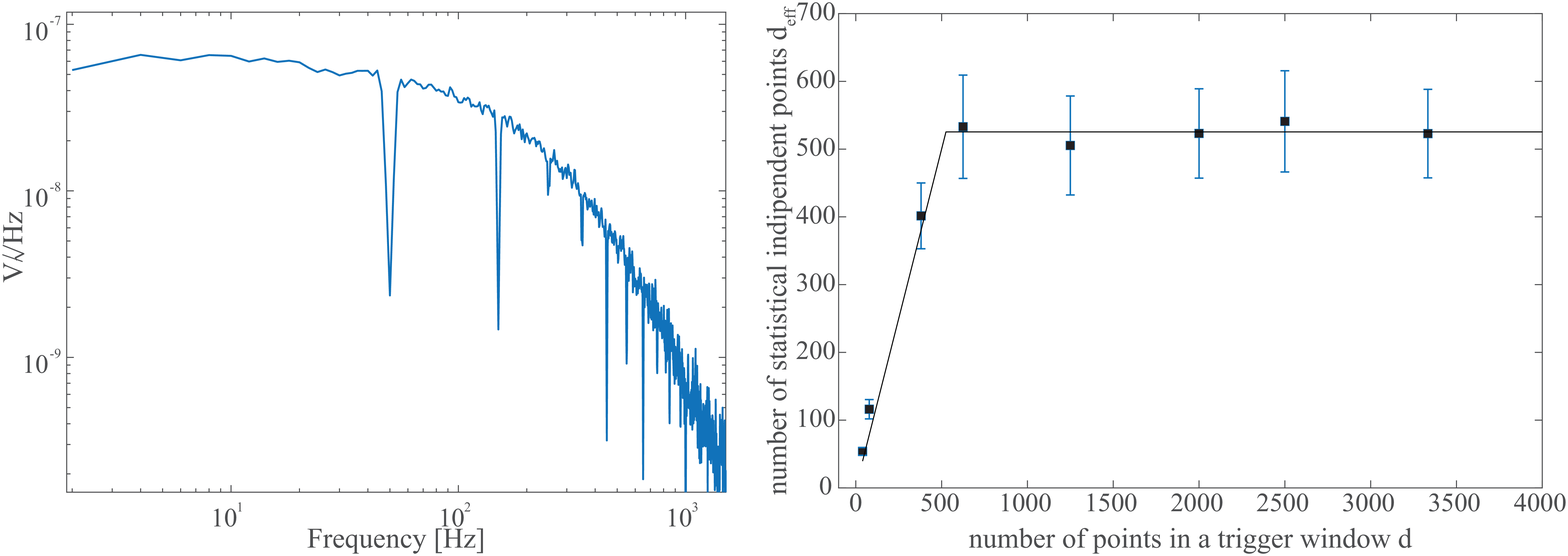}
\caption{Left: Power spectrum of filtered experimental data used in sec.~\ref{sec:exp}.  Right: $d_{eff}$ parameter as a function of the sampling rate. Each point represents the same set of waveforms of 1~s length. Varying the sampling rate results in a different number of points per waveform. As soon as the time interval between two samples is shorter than the correlation time, the $d$ parameter reaches the value of maximum number of independent samples.}
\label{fig:spec}
\end{figure}

We have discussed a method to assess the energy threshold as a function of the number of noise triggers acceptable for the experiment. We have analytically modelled the noise fluctuations under the assumption of statistically independent noise samples. During the digitalisation process already, the data are usually sampled faster than the autocorrelation time to ensure against loss of information. Moreover, the filter used introduces a time correlation among the samples and thus the value of statistically independent samples in a trigger window is reduced compared to the real number of samples in the considered time window. In other words, this filter has an effective low-pass cutting frequency $f_c$ lower than the sampling frequency. Therefore, according to the Nyquist-Shannon theorem, the time period between two statistically independent samples is $T=\frac{1}{2f_c}$. For this measurement in particular, the power spectrum of filtered baselines is shown in fig.~\ref{fig:spec} (left).\\
In case of correlated noise, where every sample follows a Gaussian distribution, the joint probability of eq.~\ref{eq:stats1} assumes the form of the general multinormal distribution:
\begin{equation}
P_{d}\left(x_{max}\right)=\frac{d}{\sqrt{2\cdot\pi}\cdot\sigma}\cdot \left(e^{-\left(\frac{x_{max}}{\sqrt{2}\sigma}\right)^2}\right)\cdot\left(\left(\frac{1}{2\pi}\right)^{\frac{d-1}{2}}\left|\Sigma\right|^{-\frac{1}{2}}\int_{-\infty}^{x_{max}}\int_{-\infty}^{x_{max}}\cdots\int_{-\infty}^{x_{max}}e^{-\frac{1}{2}\textbf{x}'\Sigma^{-1}\textbf{x}}d\textbf{x}\right),
\label{eq:stats3}
\end{equation} 
where $\left|\Sigma\right|$ denotes the determinant of the variance-covariance matrix $\Sigma$ and $\Sigma^{-1}$ is its inverse. In the general eq.~\ref{eq:stats3} we have $d-1$ nested integrals which have no analytical solution, and even a numerical approach is challenging for $d>$20~\cite{var-covar}. If the elements of the vector of noise samples $\textbf{x}$ are statistically independent, the variance-covariance matrix is diagonal and therefore the multinormal distribution can be factorized, leading to eq.~\ref{eq:stats1}.\\
To verify our interpretation of the value obtained by the fit for parameter $d$, which we understand as an effective number of statistical independent samples $d_{eff}$, we have down-sampled the filtered data at various rates and recalculated the $d$ parameter of the maxima distributions. The result is shown in fig.~\ref{fig:spec} (right). The fitted value of $d_{eff}$ stays constant where the sampling rate is above the value of the frequency corresponding to $f_c=d/t_{win}$, where $t_{win}$ is the time length of the trigger window. Below $f_c$ the value of $d_{eff}$ exactly follows the number of samples in the trigger window.\\
This shows that the lower value of $d$ can be ascribed to an oversampling of the data after filtering which does not affect the goodness of the approach, although it is numerically infeasible to predict the value of the parameter $d_{eff}$ without fitting the distribution of maxima.

\section{Conclusion and discussion}
\label{sec:end}
Many astro-particle physics experiments attempt to measure rare processes at low energies close to the energy threshold. Experiments to search for low-mass dark matter and those aimed at detecting coherent neutrino nucleus scattering, in particular, require high sensitivity for low-energy nuclear recoils in the detector. The proper choice of the (software) trigger threshold is therefore crucial. Previously this choice, which is a trade-off between the threshold and the noise trigger rate(NTR), was based largely on qualitative assumptions. We presented a quantitative method to set the energy threshold based on the expected signal rate(ESR) and the NTR.\\
For the data treatment we used the optimum filter~\cite{gatti} since it is a well-established concept to improve the energy resolution and energy threshold of a detector, and we adapted the trigger to the filtered data.\\
We have shown an analytical description of the noise trigger rate for both statistical independent and correlated noise. We have evaluated the irreducible background contribution due to triggering of baseline fluctuations with a general description applicable to many data sets. Equation~\ref{eq:stats} describes the spectral shape of this noise-induced background which can be included in future background models for better data description. We believe this is a critical step forward in rigorous data analysis close to the energy threshold.\\ 
We succeeded in evaluating the NTR and noise induced spectrum on simulated and experimental data using the method described in sec.~\ref{sec:threshold}, in good agreement with the analytical description.\\
We now consider the implication of ESR and NTR on the experimental data presented in this paper. The threshold of 19.7 eV (as in \cite{MeV}) corresponds to a NTR of 10$^6$ counts/(kg day), which considered acceptable, since triggers in noise will only contribute a fraction of 10$^{-2}$ counts of the total background rate. To give an example, we calculated the threshold for a scenario with more strict requirements: suppose this detector is deployed to investigate a physical process for which the ESR is equal to 10~counts/(kg day) in the energy interval between 20 and 40~eV and the same noise condition. A reasonable choice would be a threshold which ensures a NTR five times lower than the process under investigation. Given this expected rate, in this energy interval, the suggested threshold would be 30.3~eV (fig.~\ref{fig:maxdistreal}~(right)) which corresponds to 1~counts/(kg day). With this threshold value the misidentified triggers due to noise provide negligible contribution to the spectrum.\\
The new quantitative method proposed here is independent of the type of detector used and can be applied whenever the energy threshold is a critical parameter for data analysis.\\

\end{document}